# Scaled prototype of a tantalum target embedded in expanded graphite for antiproton production: Design, manufacturing, and testing under proton beam impacts

Claudio Torregrosa Martin,[1,2,*] Marco Calviani,[1] Antonio Perillo-Marcone,[1] Romain Ferriere,[1] Nicola Solieri,[1] Mark Butcher,[1] Lucian-Mircea Grec,[1] and Joao Canhoto Espadanal[1]

[1]*CERN, 1211 Geneva 23, Switzerland*
[2]*Universidad Politécnica de Valencia, Camino de Vera s/n, 46022 Valencia, Spain*

This study presents a further step within the ongoing R&D activities for the redesign of the CERN's Antiproton Decelerator Production Target (AD-Target). A first scaled target prototype, constituted of a sliced core made of ten Ta rods −8 mm diameter, 16 mm length-embedded in a compressed expanded graphite (EG) matrix, inserted in a 44 mm diameter Ti-6Al-4V container, has been built and tested under proton beam impacts at the CERN's HiRadMat facility, in the so called HRMT-42 experiment. This prototype has been designed following the lessons learned from previous numerical and experimental works (HRMT-27 experiment) aiming at answering the open questions left in these studies. Velocity data recorded on-line at the target periphery during the HRMT-42 experiment is presented, showing features of its dynamic response to proton beam impacts. Furthermore, x-ray and neutron tomographies of the target prototype after irradiation have been performed. These non-destructive techniques show the extensive plastic deformation of the Ta core, but suggest that the EG matrix can adapt to such deformation, which is a positive result. The neutron tomography successfully revealed the internal state of the tantalum core, showing the appearance of voids of several hundreds of micrometers, in particular in the downstream rods of the core. The possible origin of such voids is discussed while future microstructure analysis after the target opening will try to clarify their nature.

DOI: 10.1103/PhysRevAccelBeams.21.073001

## I. INTRODUCTION

Antiprotons are currently produced at CERN for the antiproton decelerator (AD) facility by impacting intense proton beams of 26 GeV/c onto a high-Z water-cooled target. The current design of the target consists in a Ir core of 3 mm diameter, 55 mm long, embedded in a graphite matrix and inserted in a Ti-6Al-4V assembly. This design configuration dates back to the late 1980s [1]. A new antiproton production target design has been foreseen for operation from 2021, aiming at improving the operation robustness and antiproton production yield for the AD physics. For this purpose, several R&D activities have been triggered during the last years, involving numerical and experimental works.

Advanced numerical tools such as hydrocodes were used to study the dynamic response of the core material each time it is impacted by the primary proton beam [2]. These simulations showed that a maximum temperature rise of 2000 °C takes place in the target core in less than 0.5 $\mu$s (the duration of a pulse burst), as a consequence of the sudden deposition of energy by the primary proton beam. This sudden rise of temperature leads to the excitation of a radial mode of vibration (also referred to as a radial wave), which exposes the material to oscillating compressive-to-tensile stresses of several gigapascals, well above its strength limit. In addition, simulations indicate that stresses in the graphite matrix may be also above its mechanical limits, due to the transmission of the generated radial wave. It is believed that this beam-induced material damage could lead to a drop of effective density of the target core, which could explain the historically observed decrease in antiproton yield during the first weeks of operation. Hence, the new design would aim at survival of the target core or, at least, reduction of its fragmentation so that the antiproton yield could be maintained.

An experiment using the CERN's HiRadMat facility [3], called HRMT-27 RodTarg, was carried out in 2015 to provide a cross-check of the numerical simulations [4,5]. The aim of the experiment was to expose different high density materials (possible candidates for the new target design) to equivalent conditions as the ones reached in the AD-Target core. This was achieved by impacting intense 440 GeV proton beams provided by the CERN's Super Proton Synchrotron (SPS) onto rods of 8 mm diameter and lengths from 140 to 240 mm (depending on the material). Targets geometries and beam parameters were selected in such a way that the mentioned rise of temperature and compressive-to-tensile stresses were also reached in the experimental targets and, at the same time, the induced vibration of the rod could be measured on its surface. For this purpose, on-line instrumentation such as a laser Doppler vibrometer (LDV) were used to record the rod surface velocity and, in this way, the performed numerical simulations could be cross-checked with experimental data. Indeed, the recorded response confirmed the presence of the predicted radial mode and its damaging consequences as well as validated the numerical simulations and strength models employed. Most of the tested high density materials (including W, Mo, TZM, and Ir among others) internally cracked even at conditions 7–5 times less demanding than the ones present in the AD-Target core (in terms of exposed rise of temperature and tensile stresses). Nevertheless, the Ta targets showed a very different response with respect to the rest of materials since, even if they presented extensive plastic deformation, they withstood the AD-Target equivalent conditions without internally cracking. This was attributed to tantalum's high ductility in comparison with the other materials tested. Therefore, Ta has become one of the main candidate core materials for the future AD-Target design.

[*]claudio.torregrosa@cern.ch

TABLE I. Comparison between properties (at room temperature) of conventional isostatic graphite (commercial grade R4550) and expanded graphite (commercial grade L30010C), both provided by the company SGL [8].

| Material | Density (g/cm$^3$) | Thermal conductivity (W/mK) In-plane | Thermal conductivity (W/mK) Through-plane | Young Modulus (GPa) | Flexural Elastic Limit (MPa) |
|---|---|---|---|---|---|
| Isostatic Graphite R4550 | 1.8 | 105 | 105 | 7 | 60 |
| Expanded Graphite L30010C | 1 | 180 | 5 | 0.5 | 1–5 |

The HRMT-27 experiment brought important insights, from a fundamental point of view, on the response of thin rods of high density materials dynamically loaded by the impact of short and intense proton pulses. Nevertheless, it left several open questions relevant for the design of an improved AD-Target. As demonstrated by means of hydrocode simulations [6], not only the structural integrity of the target core is compromised during operation, but also the one of its containing graphite matrix, especially close to the interface with the core. The fracture of the graphite matrix could have detrimental effects during operation, since appearance of cracks and gaps close to the core-matrix interface could significantly reduce the heat transfer between the core and the target cooling system (integrated in external Ti-6Al-4V assembly). Another detrimental effect of the graphite fracture could be a loss of core positioning within the matrix (taking into account that the former may be also fragmented). It is for these reasons that other possible matrix materials which could withstand the radial wave are sought. In that sense, expanded graphite (EG), also known as flexible graphite, can be a good candidate, since its flexibility could be beneficial for absorbing and damping the radial wave coming from the target core without breaking, and even adapt to its eventual changes in shape as a consequence of the plastic deformation to which is subjected. Table I shows a comparison of key properties of conventional isostatic graphite and expanded graphite, where it can be observed the potentially advantageous properties of the latter, such as higher in-plane thermal conductivity, lower density, and higher softness indicated by its low young modulus. To the best of the authors knowledge, EG has not been used as target material in any operational facility. Nevertheless, it is used as beam absorbing material in the Large Hadron Collider (LHC) External Dump (TDE, Target Dump External) [7].

Another question left open by the HRMT-27 experiment has to do with the response of Ta when subjected to a higher number of proton beam impacts since time constraints during the execution of the experiment limited the number of impacted pulses to a few.

The study presented here aims at solving these open questions introduced in the previous paragraphs. The proposed methodology consisted in the conception and manufacturing of a scaled new AD-Target prototype and its testing under proton beam impacts using the HiRadMat facility at CERN, with a new experiment named HRMT-42. The current study shows the HRMT-42 target manufacturing procedure, experiment design, on-line results, and first post irradiation examination.

## II. HRMT-42 EXPERIMENT: FIRST SCALED PROTOTYPING OF THE AD-TARGET CORE

The HRMT-42 experiment aimed at impacting around 50 high intensity proton pulses on the target presented in Fig. 1(a), which consists of ten Ta rods of 8 mm diameter, 16 mm length, embedded in a compressed EG matrix and encapsulated in a 44 mm diameter −2 mm thickness-Ti-6Al-4V container. This container was included to emulate the AD-Target envelope, to provide a body in which the EG can be compressed, as well as for safety reasons since it acted as a primary confinement to avoid radioactive contamination in case of eventual fracture of the core or matrix. The impacted 440 GeV/c proton beams had a nominal intensity of $1.5 \times 10^{12}$ ppp and a spot size of $1.5 \times 1.5$ mm$^2$ at $1\sigma$. These beams were composed of 36 bunches spaced by 25 ns (with a 5 ns bunch length), leading to an overall pulse length of 0.9 μs. Such beam parameters, as well as the core diameter (8 mm), were the same as the ones previously employed in the HRMT-27 experiment, which were specifically selected to recreate equivalent conditions (in terms of rise of temperature and radial compressive-to-tensile pressure wave) as the ones present in the AD-Target. The beam currently impacted onto the AD-Target has a momentum of 26 GeV/c, spot size of $1 \times 0.5$ mm$^2$ at $1\sigma$, and is composed by 4 bunches of 30 ns length spaced by 105 ns leading to an overall pulse length of 0.42 μs. More details of how the AD-Target conditions can be recreated in HiRadMat with the selected beam parameters can be found in Refs. [4–6].

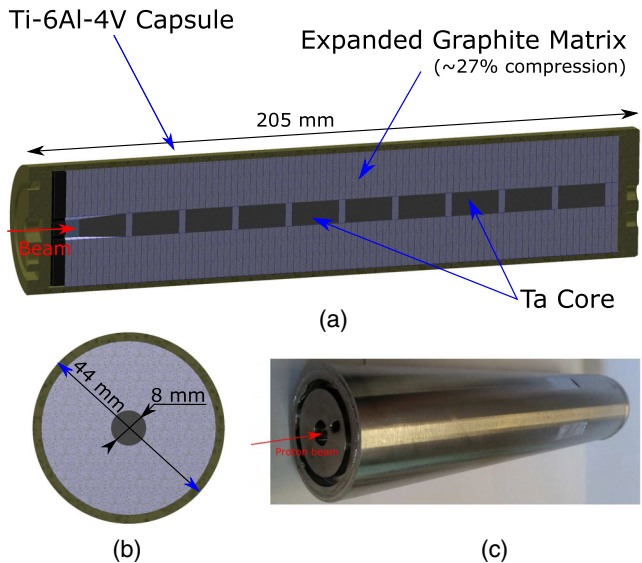

FIG. 1. (a) Half-cut of the HRMT-42 target geometry. (b) Cross-section of the HRMT-42 target geometry. (c) Picture of the target.

The HRMT-42 experiment had three main goals: (i) Assess the feasibility and manufacturing procedure of a target including compressed EG as matrix material. (ii) Study the response of EG during operation, in particular when subjected to pressure waves generated in the target core. (iii) Study the response of Ta when subjected to impact of a larger number of pulses than during the previous HRMT-27 experiment and the role of successive plastic deformations in the core, such as hardening or aging.

### A. Target geometry and manufacturing

Figure 1 shows a half-cut, a cross-section, and a picture of the HRMT-42 target respectively. The core diameter is 8 mm, similarly to the HRMT-27 experiment. This up-scaling (in comparison to the AD-Target in which the core diameter is 3 mm) is maintained to make the results of this experiment directly comparable to the ones of HRMT-27, and therefore, being able to infer conclusions during the post irradiation examination in regards to the changes in the Ta material induced by successive plastic deformation.

Nevertheless, for HRMT-42, the target core was sliced, differing from HRMT-27 in which the Ta targets were a single rod. The purpose of this slicing is to try to mitigate the excitation of bending modes, observed in that experiment, due to minor offsets in the beam impact position. The HRMT-42 target matrix is constituted by eighty seven hollow disks of EG of a thickness of 3 mm (in pristine conditions). This material, whose commercial name is SIGRAFLEX®, was purchased from the company SGL [8]. The EG disks were compressed during assembling to an average compression ratio of 27%. This compression ratio would lead to a radial expansion by Poisson effect estimated in the order of 150 μm (in unconstrained conditions), which was deemed enough to fill the potential gaps between the Ta-EG and EG-Titanium due to tolerances, as well as to apply certain contact pressure to guarantee thermal conductance, or adapt to potential swelling of Ta core due to plastic deformation. Seventeen small disks of 8 mm diameter of EG were also added in the central region, in between the Ta rods. These small disks compensate the incompressibility of the region of the core where the Ta is present. The upstream tantalum rod had a conical shape to facilitate the insertion of hollow disks during compression, since a relative movement between the central and periphery regions of the matrix takes place, as illustrated by Fig. 2.

Figure 2 shows the assembly procedure of the target, for which dedicated compression and welding tooling were designed, built and employed. In addition, two half-tubular pieces were inserted on top of the target to guide in the EG during the compression. Once the compression was done, these tubular pieces could be removed so the electron beam could have vertical access to weld the lid. During the welding phase, the pressure on the matrix was kept by two locking nuts. In addition, the welding over the whole edge of the lid could be carried out in just one turn, since the tooling was equipped with bearings which allowed the target rotation on its axis.

Figure 3 shows a picture of the different pieces of the target before assembly. Cu foils of 0.1 mm thickness in form of hollow disks were inserted between some of the EG in order to make their interface visible to x-rays and, in this way, track the uniformity of the EG compression over the target's length. The maximum pressure applied during compression of the EG matrix was around 30 MPa, which achieved an average compression ratio of 27%. The elastic recovery associated to this compression was measured to be in the order of 5%. However, uniform compression was not achieved. This can be seen in Fig. 4, which shows an x-ray image of the target before irradiation. The position of the Ta core as well as the Cu foils within the EG matrix are revealed by the image. There is a larger separation of the Cu foils at the bottom of the target than at the top, indicating that the compression ratio was ranging from only 7% at the bottom, up to 40% in the top. This uneven compression is consequence of the friction forces of the EG with the capsule walls, which were preventing the transfer of pressure to the EG layers of the bottom. A possible way to avoid this issue in the future design could be by inserting and compressing the EG matrix gradually, instead of compressing the whole stack at once.

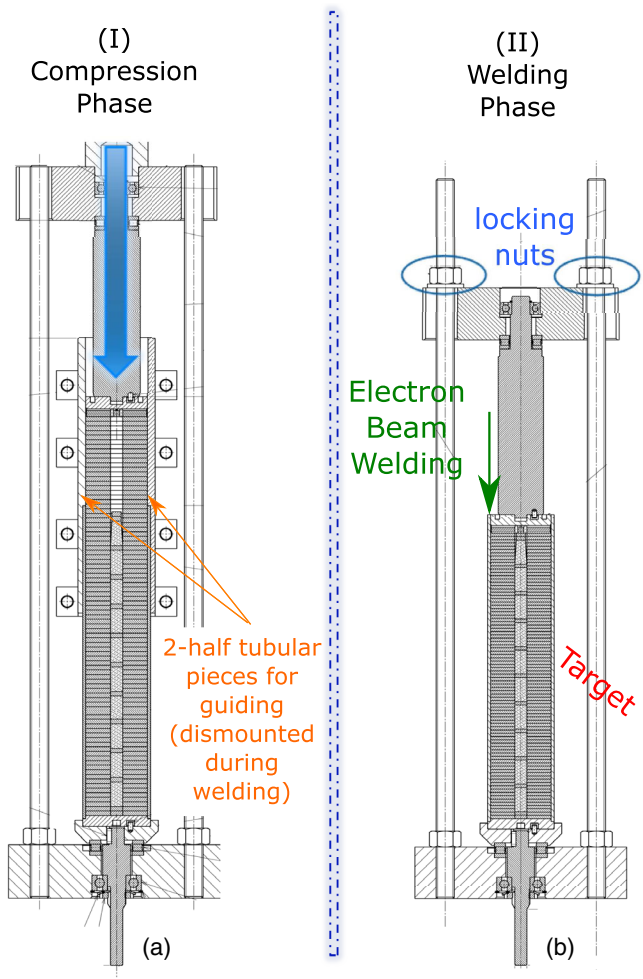

FIG. 2. (a) HRMT-42 target assembly before compression. (b) Target assembly with the EG-matrix compressed, ready for EBW.

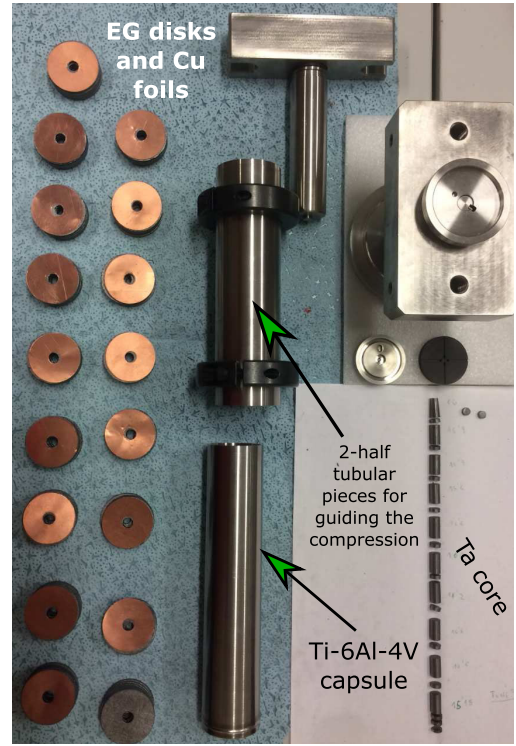

FIG. 3. Picture of the different parts of the HRMT-42 target and auxiliary pieces before assembly.

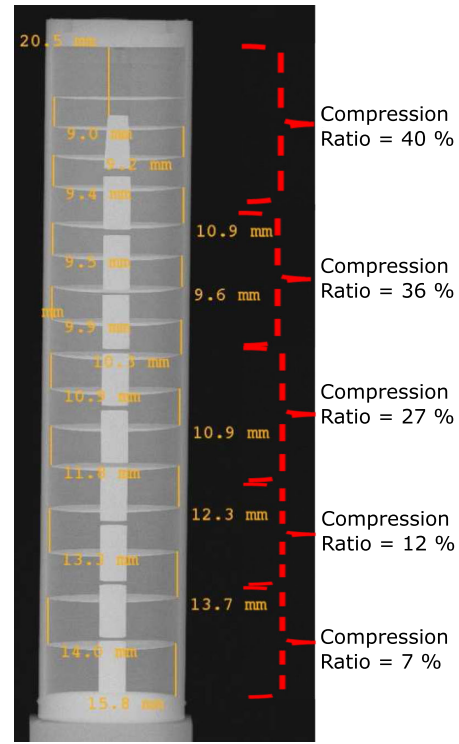

FIG. 4. X-ray image of the target before irradiation showing the Ta rods of the core and Cu foils within the EG matrix.

## B. The HRMT-42 experiment setup and conditions in the target

The HRMT-42 target was placed in a specially designed sample holder (Fig. 5), located inside a stainless steel tank filled with air, with a slight underpressure with respect to the atmosphere achieved by a aspiration system with HEPA filters at the exhaust vent line. The tank and experimental setup were reused from the previous HRMT-27 experiment. It consisted of experimental tables providing horizontal and vertical motorized movements of the tank, used for alignment purposes. The experimental tables were mounted on the standard HiRadMat table, which provides a plug-in interface for a quick installation in the beam line.

Inside the tank, the sample holder could move vertically driven by a step motor and a LVDT (linear variable differential transformer), allowing the positioning of each of the corresponding samples in the proton beam trajectory. The clamping system of the HRMT-42 target to the sample holder was specially designed in order to allow for fast and remote (robot friendly) extraction of the target after the experiment, thus reducing the exposure of personnel to ionizing radiation. This system relied on a single M12 screw which, when unscrewed, released the target by gravity to a position in which it could be grabbed by a robot.

In addition to the HRMT-42 targets, the sample-holder was equipped with a 100 mm long CuCr1Zr mask with a rectangular aperture of $4 \times 8$ mm. This mask was used for beam-based-alignment of the experimental setup. Beam size and pulse-to-pulse beam stability were monitored by a beam TV screen [9] working in vacuum, which is placed upstream, in the experimental table A of the HiRadMat line.

In order to record the dynamic response of the HRMT-42 target, the experiment counted on a Laser Doppler Vibrometer (LDV) with a purely passive optical head placed inside the tank, pointing at a flat surface machined on the Ti-6Al-4V capsule of the target. This LDV had an acquisition rate of 4 MHz. Outside the tank, pointing through a radiation hard glass, two cameras were used to monitor the experiment; a conventional high definition camera and a rad-hard camera.

The characteristics of the proton beam impacted on the target were the same as the highest intensity ones of the HRMT-27 experiment [4,5] which were already introduced in the first paragraph of Sec. II. These parameters were selected in such a way as to excite the radial vibration mode of the core, subjecting it to the compressive-to-tensile pressure states taking place in the AD-Target [2]. Figure 6 shows the estimated temperature profile in a quarter of the target's geometry as a consequence of a proton pulse impact. This temperature profile was obtained based on depositions of energy by the beam simulated by means of FLUKA Monte Carlo simulations [10], one-way coupled with the Finite Element solver of ANSYS Mechanical. Based on these simulations, a maximum temperature of 1850 °C is reached in the fourth upstream Ta rod.

Figure 7 shows the pressure response at the center of core rods 2, 4, 7 and 10 estimated by AUTODYN® simulation following a similar methodology as in Ref. [2]. As anticipated, a compressive-to-tensile pressure wave with a period of 2.7 μs (corresponding to a radial mode of vibration) dominates the response, reaching up to 5 GPa in tension. This simulation considered a Mie-Grüneisen [11] Equation of State and a Johnson-Cook strength model obtained from Ref. [12]. This strength model takes into account the response beyond yielding as well as the temperature and strain rate dependence of the material. Nevertheless, this simulation corresponds only to the first impacted pulse, since the change of material properties or fracture due to successive pulses cannot be easily taken into account by the simulation. The following section presents the results of the on-line monitoring.

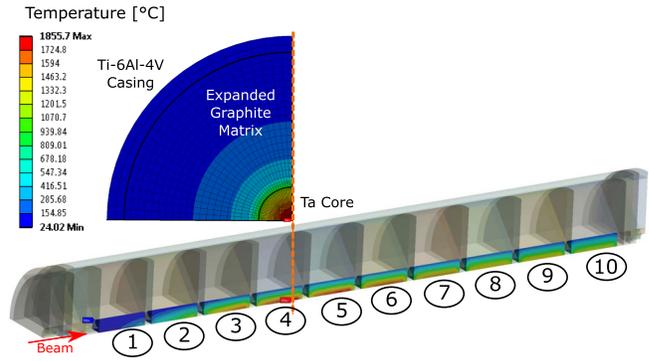

FIG. 6. Contour plot showing the temperature reached in the HRMT-42 target as a consequence of a proton beam impact.

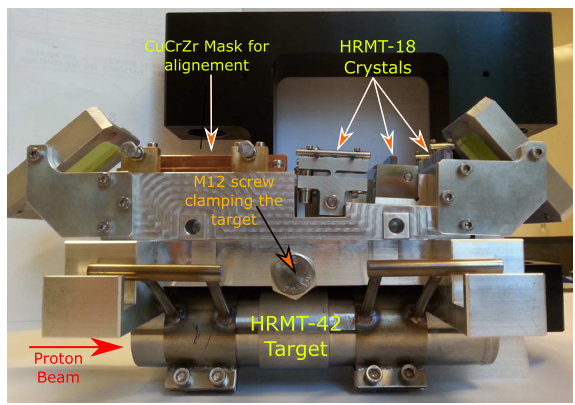

FIG. 5. Picture of the sample holder used in the experiment, which was carrying the crystal samples (irradiated within the experiment HRMT-18), and the HRMT-42 target.

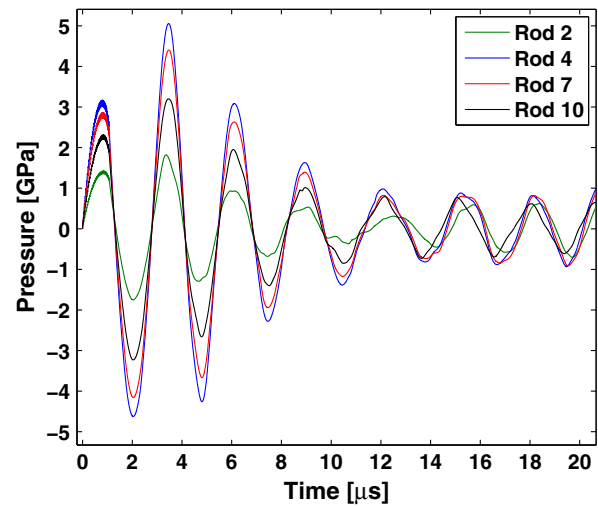

FIG. 7. Pressure response at the center of core rods No. 2, No. 4, No. 7 and No. 10 when impacted by the HiRadMat beam, estimated by hydrocode simulations using the software AUTODYN®.

### C. Operation and on-line results

A total of 47 high intensity proton pulses were impacted on target during the experiment, leading to a total POT (protons on target) of $7 \times 10^{13}$. The repeatability of the prompt radiation measurements between successive pulses measured by the radiation monitors in the area was already suggesting on-line that no bending of the Ta core was occurring, in contrast to what happened in the HRMT-27 experiment. In the latter, a progressive decrease of the peak prompt radiation was observed, which was attributed to successive bending of the Ta target and subsequent loss of beam target-interaction.

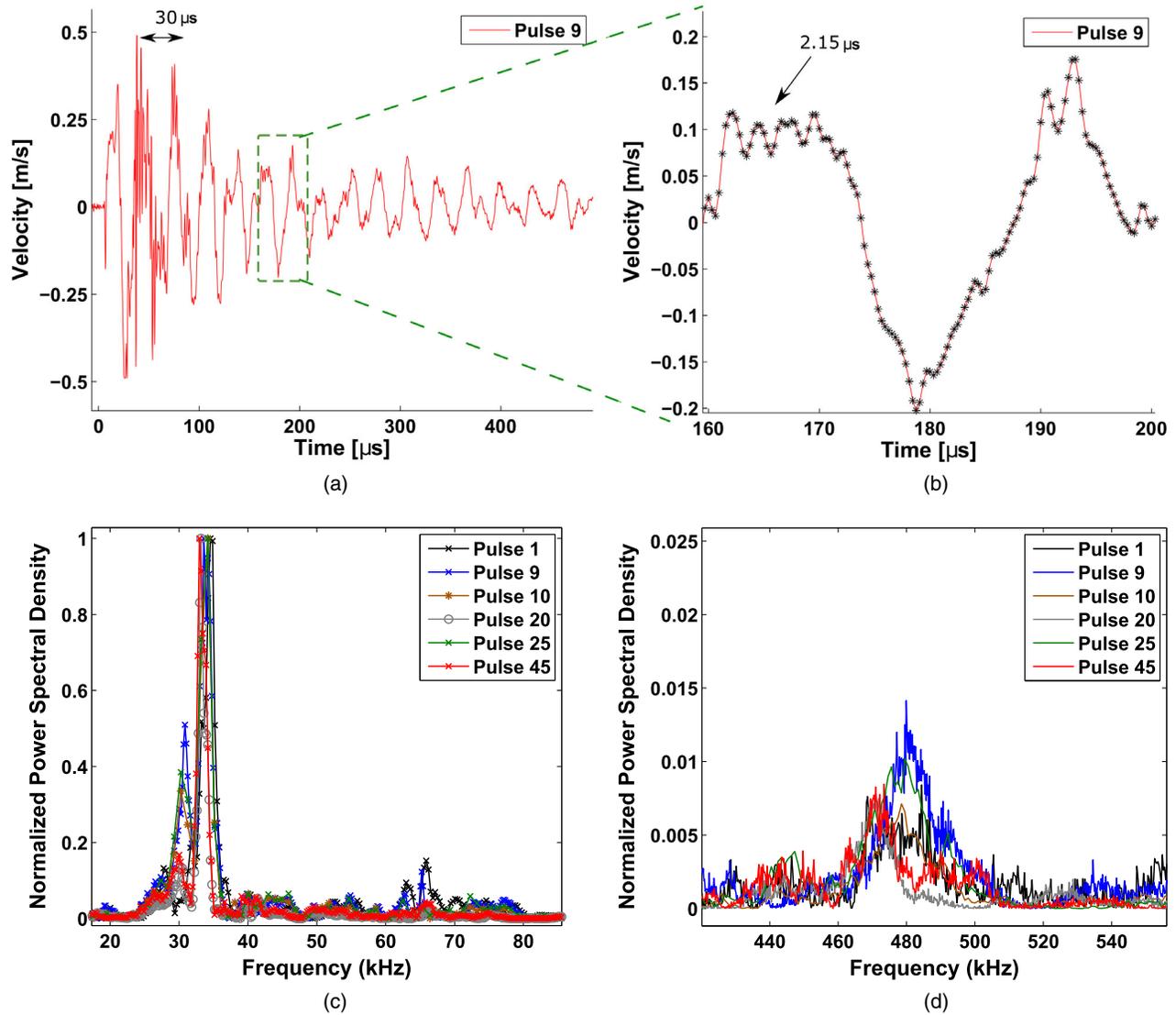

FIG. 8. (a) Example of radial velocity recorded at the HRMT-42 target periphery by the LDV during the ninth impacted pulse. (b) Zoom-in of the recorded velocity showing the high frequency component of the recorded velocity. This is attributed to the radial mode of vibration of the Ta core which propagates towards the target periphery through the EG matrix. (c) Frequency domain of the velocity recorded by the LDV for different impacted pulses, showing the low frequency range. Components within 30–35 kHz identified as radial modes of vibrations of the whole target dominate the velocity response. (d) Zoom-in of the frequency domain of the recorded velocity showing the high frequency range. A peak within 470–480 kHz can be observed which is identified as the radial mode of vibration of the Ta core.

Figure 8(a) shows the velocity measured by the LDV for the ninth impacted pulse. The flat surface at which the LDV pointed was located at 75 mm from the downstream part of the target, which corresponds to the zone in which core rod number 7 is placed (as defined in Fig. 6). The maximum velocity was in the order of 0.5 m/s. A wave with a period in the order of 30 $\mu$s dominates the velocity response. This period corresponds to the radial mode of vibration of the whole target. The estimated speed of propagation of this wave within the matrix is in the order of 1300 m/s, which is considerably slow if one compares it, for instance, to the bulk speed of sound in conventional isostatic graphite, which can be in the order of 3000 m/s. This slow wave propagation is attributed to damping and to the porous nature of EG matrix. The damping of the recorded velocity takes place within 500–600 $\mu$s as shown in Fig. 8(a).

Figure 8(c) shows a Fourier transformation of the recorded velocity during several pulse impacts. As introduced, the dominant components of the response are within 30–36 kHz. The figure also shows a relatively good repeatability in the frequency domain between different impacted pulses, even though the high amplitude of the generated pressure wave led to significant changes in the core geometry and matrix in each impacted pulse (as it will be seen in the topographies shown in the next section).

Looking in detail at the recorded velocity, it is also possible to observe a higher frequency component. This is shown for example in Fig. 8(b), which corresponds to a zoomed-in region of the 160–200 $\mu$s time window of Fig. 8(a). As can be seen in the figure, a ∼2.15 $\mu$s period oscillation is present in the velocity response. Such a high frequency can only correspond to the radial mode of vibration of the Ta core of the target, which oscillations are propagated through the EG matrix, reaching the Ti-6Al-4V external envelope. These components can be also observed in Fig. 8(d), which shows the high frequency region of the Fourier transformation of the velocity response for several pulses. A peak in the order of 470–490 kHz can be observed. The repeatability between pulses is not as remarkable as in the lower frequency domain. This could be due to progressive changes in the Ta core due to the successive plastic deformations taking place as a consequence of each proton beam impact. Nevertheless, a good indicator of the contact behavior between the Ta core and the EG is the fact that this high frequency component is appreciable at the target's surface even after several hundreds oscillations and many pulses. This was one of the main motivations of the test. Another good indicator of the good performance of the EG matrix is the high repeatability in amplitude and damping of the velocity recorded over the 47 proton pulses impacted. This would suggest that the contact within the Ta core and EG is continuous since the induced waves propagated similarly towards the target periphery even if the core was suffering significant plastic deformation with every pulse.

Finally, it is worth highlighting an unpredicted phenomenon concerning the high frequency component in the measured velocity, shown with a ∼2.15 $\mu$s period in Fig. 8(b) and a 470–490 kHz frequency peak in the Fourier transformation of Fig. 8(d). This component is attributed to the radial mode of vibration of the Ta core since there is not any other body within the target geometry that can vibrate in such a high frequency range. However, both in the hydrocode simulations presented in Fig. 7 and in the experimental data recorded for 8 mm diameter Ta targets in the previous HRMT-27 experiment [4,5], as well as in analytical predictions of elastic radial modes presented in Ref. [6], the period of such mode was ∼2.7 $\mu$s instead of ∼2.15 $\mu$s. Therefore, the mode measured in the present experiment is 25% faster than that in the simulations and in the measurement of the previous experiment. It is certainly not straightforward to explain this difference since specific effects of HRMT-42 such as temperature softening would lead to the opposite effect. Potential coupling between radial and longitudinal modes due to shorter length of the Ta rods (16 mm length in comparison to 160 mm the HRMT-27 experiment) is not the reason either, since such frequency shift would have been observed in the AUTODYN® simulations presented in Fig. 7.

Another possible explanation conceived would have been that the internal fracture of the Ta core (suggested by the preliminary PIEs shown in next section) reduces the effective material density within its radius, leading to a higher frequency vibration. Nevertheless, one would expect that such fracturing would distort this radial wave, while Fig. 8(b) showed a relatively clean wave even after several pulses and hundreds of thousands oscillations per pulse. In addition, if this would have been the case, a significant frequency shift should have been observed from pulse 1 to pulse 45, which did not take place. The only explanation left would be that this effect is related to the Ta hardening due to extensive plastic deformation, which takes place from the first pulse, with a possible change in the effective Young modulus. Future post irradiation examinations after target opening and core extraction, such as hardness measurements, will be performed to try to clarify this phenomenon.

### III. FIRST PIES OF THE HRMT-42 EXPERIMENT

In addition to the on-line data presented in the previous subsection, the experiment strongly relied on the post irradiation examinations to evaluate the performance of the Ta core and EG matrix, in particular at the contact interface between them. Even if the on-line data can be very useful to identify dynamic phenomena taking place as a consequence of each proton beam impact, it is somehow limited since the point of velocity measurement was at the target periphery and far from the core. The challenges of the post irradiation examinations reside in being able to study the Ta-EG interface without modifying it during the target opening process, i.e., to find the proper nondestructive testing. In that sense, it was found that the x-ray and neutron tomographies can be very powerful techniques to achieve this goal. Nevertheless, all analyses are complicated by the residual dose rate of the target assembly, which was measured at 400 $\mu$Sv/h at contact 20 days after the irradiation.

### A. X-ray tomography at the ESRF

The European Synchrotron Radiation Facility (ESRF) in Grenoble can provide high energy x-rays to multiple experimental lines. For the inspection of the HRMT-42 target, the beamline ID9 [13] was used, which delivered 160 keV photons, achieving a spatial resolution of ∼12 $\mu$m. This high energy and spatial resolution was necessary in order to optimize the inspection of the Ta-EG interface. Conventional x-rays are limited in doing so due to the huge difference in density between these two materials, which would produce a blurry halo at their interface.

In the inspection, the target was placed vertically in a motorized stage, which could rotate within the axis of the target to achieve a 360° imaging, required for performing the tomography. Five zones of the target were inspected over an 8 mm length in each of them. The inspections at the central and downstream zones of the target were prioritized, since higher temperature and dynamic stresses where reached in these parts. The target was inspected before and after the proton irradiation at the HiRadMat facility, so the induced changes could be directly comparable.

Figure 9 shows x-ray images of Ta rod No. 4 and No. 6 respectively inside the EG matrix, before and after irradiation in HiRadMat. The Cu foils used for tracking the compression are also observable. These rods were within the most loaded ones, subjected to estimated adiabatic rises of temperature of 1850 °C and tensile pressures close to 5 GPa. The images clearly show a considerable plastic deformation of the Ta core. This plastic deformation includes mainly a slight bending of the rod (most probably caused by excitation of a bending mode due to minor offsets in the impact position with respect to the axis of the rod), as well as protuberances in the order of 0.5 mm at the trajectory of the beam, within a radius of 1 mm. These protuberances, together with a displacement of the base of the Ta rods due to bending, lead to the appearance of gaps between the rods and the EG internal disks with a size within 1 mm. In addition, it seems that plastic deformation of the Ta core led to 0.5 mm swelling in the radial direction. Nevertheless, the image shows that the EG matrix adapts to such change in shape of the core and no gaps appear there. This is a remarkable result since it would mean that the thermal contact is guaranteed and heat could be properly evacuated from the target core in the new design. On the other hand, 160 keV x-rays are not powerful enough to penetrate the high density Ta core and they are unable to evaluate its internal structural state. For this reason, another nondestructive technique based on neutron imaging was carried out to complete this study.

### B. Neutron tomography at Neutra Line at PSI

The Neutra line (for NEUtron Transmission RAdiography) [14] at the Paul Scherrer Institute (PSI) in Switzerland counts on an almost parallel beam of thermal neutrons coming from the Swiss Spallation Neutron Source (SINQ). This line is specialized in neutron radiography and tomography images of medium to large size objects using a thermal neutron energy spectrum, which is convenient to penetrate through the 8 mm diameter Ta core (its attenuation coefficient for thermal neutrons is around 17 times lower than that for 150 keV x-rays). The Neutra line also has the advantage of being able to accept radioactive samples. Similarly to ESRF, a neutron tomography of the HRMT-42 target was carried out. Nevertheless, the large aperture of the neutron beam allowed to scan the whole target length by just dividing it in two scanning zones, upstream and downstream. The experimental spatial resolution was in the order of 50 $\mu$m.

Figure 10 shows a colored contrast image corresponding to a longitudinal cut of the performed tomography. The image clearly shows the ten rods of the Ta core slightly bent and deformed and, as opposed to the x-ray imaging, it reveals their internal structure. As can be observed, most of the Ta rods show internal discontinuities with lower density regions which can be interpreted as the presence of voids. These voids appear in different planes and, differently from what one could expect, are not necessarily within the axis of the rods (where higher tensile pressures are reached according to simulations) but circumferentially distributed over a 1–2 mm radius. This can be more easily appreciated in Figs. 11 and 12, which show transversal cuts of rods No. 4 and No. 9 respectively. The dimensions of these voids range from some tens of micrometers in radius (limited by the resolution) by 500 $\mu$m in length (rod No. 4 in Fig. 11), up to 800 $\mu$m in radius by ∼1.8 mm in length (rod No. 9 in works found in the literature studying this phenomenon are based on flyer impact tests in plates, while in the current application the tensile load comes from the proton beam impact and the associated temperature rise, which may play and additional role. It is also interesting the fact that the Ta rod of the HRMT-27 experiment (impacted only a few times) did not show these voids, suggesting that subsequent yielding may have played an additional role. Furthermore, the rest of the refractory metals tested in HRMT-27 were substantially more brittle, which could explain the different mode of fracture observed for them, mainly characterized by longitudinal cracks without this void formation.

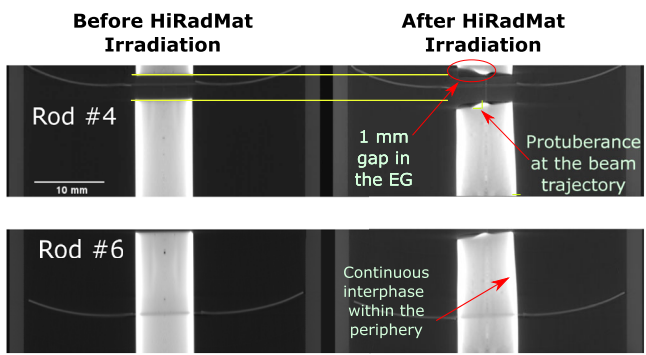

FIG. 9. Images of rods No. 4 and No. 6 of the Ta core obtained from x-ray tomographies performed at ESRF. These tomographies were performed before and after proton irradiation in HiRadMat. Extensive plastic deformation in the Ta core induced by the successive proton beam impacts can be appreciated.

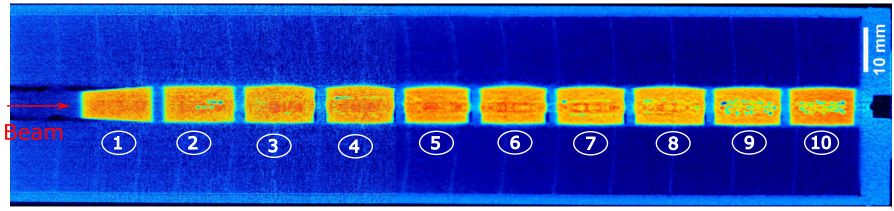

FIG. 10. Colored contrast image of the neutron tomography performed in the HRMT-42 target after proton irradiation. The images reveals the presence of voids in the Ta core, particularity concentrated within the downstream rods. Note that this image is obtained by merging two images corresponding to the tomographies of the upstream and downstream part. Therefore, the image contrast at the matrix between these two areas is not directly comparable.

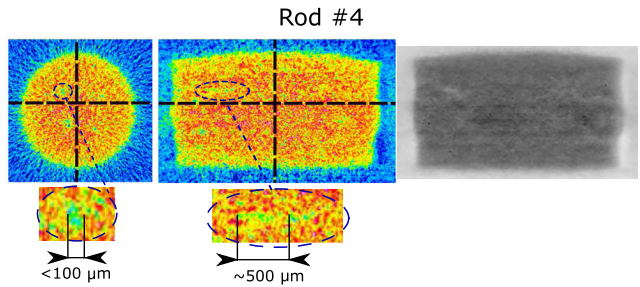

FIG. 11. Zoom-in images of the neutron tomography corresponding to rod No. 4 (the most loaded one according to simulations presented in Figs. 6 and 7). Voids with a size of a few tens of micrometers in radius by 500 $\mu$m can be identified.

Fig. 12). The fact that these voids do not seem to join to form longitudinal cracks as it was observed with the other refractory metals irradiated during the HRMT-27 experiment [6] is remarkable. The origin of these voids may be strongly related to so called, spall fracture, which is a phenomenon commonly reported in the literature describing the behavior of ductile metals, such as tantalum, subjected to high dynamic loading [15,16]. This fracture is characterized by nucleation, growth, and finally coalescence of micro-voids induced by intense tensile pressures. Some studies mention a tensile pressure threshold of 3 GPa above which these voids can nucleate in tantalum [16,17]. As suggested by Fig. 7, these tensile pressure were exceeded. Nevertheless, it shall be mentioned that the

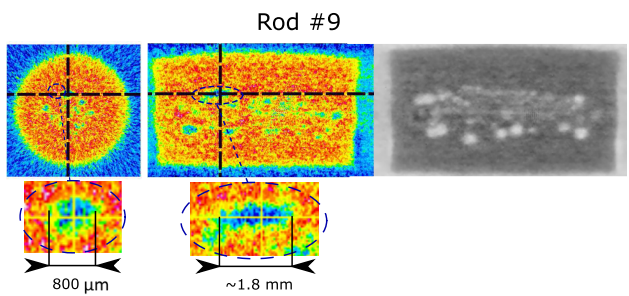

FIG. 12. Zoom-in images of the neutron tomography corresponding to rod No. 9, which apparently is the most damaged one. Voids with a size of $\sim$800 $\mu$m in radius by $\sim$1.8 mm in length can be identified.

Another somehow surprising feature observed in the image of Fig. 10, and by comparing Figs. 11 and 12, is that most damaged rods are not necessarily the most loaded ones according to the simulations. Figures 6 and 7 showed that the higher temperatures and tensile pressures were reached in rod No. 4, while the neutron tomography suggests that the most damaged ones are the downstream ones, No. 9 and No. 10. It must be mentioned in any case the energy load in these last rods was not negligible, leading to an adiabatic rise of temperature in the order of 1300 °C and tensile pressures of $-3$ GPa. It shall be also mentioned that these are preliminary results and subsequent PIEs consisting in micro-structure analysis of the core after target opening are necessary for a full understanding of the extent and causes of Ta damage. In any case, it is interesting that the apparently most damaged rods are the ones which were exposed to temperatures around the Ta recrystallization temperature (1200–1400 °C), while the rod No. 4 exceeded these temperatures, being potentially subjected to recrystallization. Further detailed PIEs could bring new insights to these phenomena. It was generally believed that the material was not exposed for a sufficiently long time to high temperatures to activate such recrystallization processes, but it may not be so.

## IV. CONCLUSIONS

The current study is another important step in the framework of the ongoing R&D activities for the design of the new CERN AD-Target, which started in late 2013. These activities have been mainly focused on reducing the existing uncertainties regarding the dynamic response of the AD-Target core and its containing matrix due to the sudden energy deposition by primary proton beam. In this context, the previous studies were dedicated to the response of the high density core, by means of both numerical simulations (using hydrocodes [2]) and experimental works using the CERN HiRadMat facility (HRMT-27 experiment [4,5]). These works brought important findings regarding the dynamic phenomena taking place in the target core, and that the response of Ta was significantly better than other high density candidate materials (such as Ir, W, Mo and TZM) thanks to its high ductility. However, they were fundamental studies which lacked conclusions applicable to a new target manufacturing and the long term response of Ta when subjected to a large number of proton beam impacts. The manufacturing of a first scaled prototype—constituted of ten 8 mm diameter Ta rods embedded in a EG matrix inside a Ti-6Al-4V container—and its testing within the HRMT-42 experiment, aimed at filling this gap.

The present study has proposed and executed an innovative concept of target core matrix, by using a compressed EG instead of conventional isostatic graphite, and the feasibility of its assembling within a Ti-6Al-4V electron beam welded container.

The HRMT-42 experiment successfully impacted 47 pulses in the manufactured prototype core, bringing it to equivalent conditions to the ones present during AD-Target operation. The recorded on-line data (consisting of target velocity at its periphery) showed that the velocity response is dominated by a ∼30 $\mu$s wave, identified as the radial mode of the whole target. This data also showed the high damping properties of EG, attenuating this wave within only 500–600 $\mu$s. Even if it is not in the scope of this work, this velocity data could be potentially used for modeling EG and benchmarking its dynamic properties, which is not a straight forward task due to the high anisotropy of the material. These properties can be valuable for other beam intercepting devices at CERN such as the Large Hadron Collider external dump. The recorded velocity also showed a high frequency component, with a period of 2.1 $\mu$s identified as the radial vibration of the Ta core, propagating towards the EG matrix until the periphery. However, this period is slightly lower than the one observed for 8 mm diameter Ta rods in the previous HRMT-27 experiment as well as in simulations. The reason behind this phenomenon is not well understood yet.

This study has shown innovative forms of non destructive PIEs applied to HiRadMat targets, such as x-ray and neutron tomographies. The high resolution of the x-ray tomography revealed the extensive plastic deformation of the Ta core and the performance of the Ta-EG interface, while the neutron tomography could reveal the internal state of the Ta core. The x-ray tomography suggests that the EG matrix is able to adapt to the change of shape of the core due the plastic deformation, in particular through its periphery. This is a positive result since it would mean that the thermal contact is guaranteed and heat could be evacuated through the EG disks, which show the highest thermal conductivity within their in-plane direction. This tomography also shows the formation of ∼1 mm gaps between the Ta rods and internal 8 mm diameter disks of EG, due to a plastic protuberance appearing in the former. These gaps are intrinsic to the plastic deformation of the Ta and it seems difficult to avoid their formation. Nevertheless, their contribution to the heat removal of the core is minimal, since heat transfer will take place mainly at the periphery.

Finally, the performed neutron tomography revealed the formation of voids inside the Ta core. This phenomenon was not observed during HRMT-27 in which Ta rods were exposed to the same conditions but just for a few high intensity pulses. Future microstructure analysis foreseen after opening the target will try to clarify the origin of such voids, which seem to be remarkably similar to the ones described in the literature specifically addressing spalling fracture in Ta [16]. In any case, the main conclusion applicable to the future target design is that an additional effort to reduce the tensile stresses in the target core should be made, either by increasing the core diameter or/and by increasing the beam size. Nevertheless, tantalum is still considered to be one of the main candidate materials since its behavior is still better than the other refractory metals available, which fractured just after a single proton pulse impact during the HRMT-27 experiment.

This paper was focused on the dynamic response of the target prototype and did not assess other operation aspects such as radiation damage and its influence in the detrimental of mechanical and thermal properties of the Ta core and EG matrix. Simulations performed by FLUKA Monte Carlo simulations estimate that up to 1 dpa/year would take place in the Ta core during AD-Target operation. The low total POT provided by HiRadMat prevents the use of this irradiation to investigate radiation damage. Nevertheless, it exists valuable data in the literature of post irradiation studies of Ta exposed to 800 MeV proton irradiation up to 11 dpa in the ISIS spallation target, showing that Ta retained very high ductility after proton irradiation even at the highest doses [18]. This result suggests that the hardening due to successive Ta plastic deformation during AD-Target operation would play a more important role than the one produced by radiation damage. Regarding the radiation damage on the EG matrix and, in particular, the degradation of its thermal conductivity, a 180 MeV proton irradiation campaign including this material (up to 0.04 dpa) was carried out during 2017 using the BLIP facility at Brookhaven National Laboratory (BNL) in the context of the RaDIATE International Collaboration [19]. The PIEs results of this irradiation will help to assess this effect.

In addition, the manufacturing of new AD-Target prototypes and preparation of its testing in HiRadMat is currently ongoing, with the so called PROTAD experiment, which will take place in 2018 as the last step of the AD-Target redesign process. These prototypes will be real size, with much smaller core diameter than the one presented in this study. This smaller diameter will be compensated by a more focused and shorter HiRadMat proton beam, but with a lower intensity to also recreate in this way AD-Target conditions. Six different prototypes will be manufactured, including matrices made both of EG and conventional isostatic graphite (so their response can be compared) as well as different core configurations with rods with diameters larger than 3 mm. In addition, all of these prototypes will count on the external air-cooled envelope, as in the real AD-Target operation. This air cooling system aims at replacing the current water-cooled design, simplifying the operation in terms of robustness and cooling fluids activation management.

## ACKNOWLEDGMENTS

The authors want to express their gratitude to the different people involved in the manufacturing of this prototype, in particular to E. Grenier-Boley, D. Grenier, and K. Kershaw for supporting the design, to M. Crouvizier during the compression tests, to L. Prever-Loiri for the EBW of the Ti-6Al-4V capsule, and to G. Arnau and A. Porret for the preliminary x-ray inspections performed at CERN. In addition, the authors want to thank the HiRadMat Facility team as well as all CERN groups involved in the execution of the experiment, such as BE-OP-SPS, BE-BI-PM and HSE-RP-AS. Furthermore, special thanks to E. Boller and P. Tafforeau from ESRF for their support in the x-ray tomographies as well as C. Grunzweig, D. Mannes and M. Strobl from PSI for the neutron tomographies. Finally, the authors want to thank the CERN's Accelerator Consolidation (ACC-CONS) project, which financed this work. In addition, the research leading to these results has received funding from the European Commission under the FP7 Research Infrastructures project EUCARD-2, Grant Agreement No. 312453. Lastly, the authors are grateful to Anna Lambert for the English proofreading of this manuscript.